\documentclass[12pt]{revtex4}
\begin{document}
\title{Dispersed Phase of Non-Isothermal Particles in Rotating Turbulent Flows}

\author{R.~V.~R.~Pandya}
\email[]{rvrptur@yahoo.com}
\author{P.~Stansell}
\email[]{paulstansell@gmail.com}

\date{\today}

\begin{abstract}
We suggest certain effects, caused by interaction between rotation and
gravitation with turbulence structure, for the cooling/heating of
dispersed phase of non-isothermal particles in rotating turbulent
fluid flows. These effects are obtained through the derivation of
kinetic or probability density function based macroscopic
equations for the particles. In doing so, for one-way temperature
coupling, we also show that homogeneous, isotropic non-isothermal
fluid turbulence does not influence the mean temperature (though
it influences mean velocity) of the dispersed phase of particles
settling due to gravitational force in the isotropic turbulence.
\end{abstract}

\narrowtext

\maketitle

\section{Introduction}
Different phenomena of non-isothermal two-phase flows existing in
situations, such as, dispersion of aerosols, dust devils, dust storms
and their effects on atmosphere, protoplanetary nebula and disks,
birth and growth of cloud droplets, all involve rotation,
gravitational field (e.g. see
\cite{RY89,Friedlander00,Bannon02,GL04,KRGB04}). Researchers continue
to face challenges in accurately explaining the observed and
discovering the new phenomena
\cite{CTTV75,EF94,EKR98,BFF01,PM02b,PSC04} due to the involved
turbulence closure problem with complexity further enhanced by the
presence of dispersed phase of particles and droplets. In recent
years, combination of kinetic approach \cite{Boltzmann64} and theories
for classical fluid turbulence closure problem \cite{McComb90} has
evolved in a mathematical rigorous approach of probability density
function (PDF) capable of capturing certain phenomena for the
dispersed phase of particles/droplets in turbulent flows (e.g. see
reference \cite{PSC04} and references cited therein). In our ongoing
efforts toward unification of various aspects and explanation of the
phenomena of two-phase turbulence \cite{PSC04,PM03,PM02b}, in this
letter we apply PDF approach
\cite{DZ90,Reeks91,Reeks92,Zaichik99,PM99,Derevich00,MP01} to
important situation of two-phase rotating, non-isothermal, turbulent
flows yielding new effects for collective behavior of particles'
temperature.

\section{Analysis}
Consider particles dispersed
in non-isothermal rotating flows moving due to forces of fluid drag,
rotation (Coriolis and centrifugal) and gravity. Particles also
exchange heat with the surrounding fluid and are considered here as
point particles for analysis purpose. The trajectory and temperature
of each particle is governed by the Lagrangian equations for its
position ${\bf X}$, velocity ${\bf V}$ and temperature $T_p$,
written in a coordinate system having constant angular velocity
${\bf \Omega}$ as
\begin{equation}
\frac{dX_i}{dt}=V_i, \quad
\frac{dV_i}{dt}=\beta_v(U_i-V_i)+2\epsilon_{iab}V_a\Omega_b +f_i,
\label{l1}
\end{equation}
\begin{equation}
\frac{dT_p}{dt}=\beta_\theta (T-T_p)+Q(T_p). \label{l2}
\end{equation}
Here, $\beta_v=1/\tau_p$, $\tau_p$ is particle velocity time
constant, $U_i$ and $T$ are carrier fluid velocity and temperature
in the vicinity of the particle, $\epsilon_{iab}$ is the
Levi-Civita's alternating tensor, and $f_i=f_i^c+f_i^g$ account for
components of centrifugal ${\bf f}^c=-{\bf \Omega}\times ({\bf
\Omega} \times {\bf r})$ and gravitational ${\bf f}^g=f^g{\bf r}$
accelerations with $f^g =-(GM/r^3)$ and position vector of particle
${\bf r}$ and $r=\mid {\bf r}\mid$. Also, $G$ is the gravitational
constant, $M$ is the mass responsible for ${\bf f}^g$,
$\beta_\theta$ is inverse of the particle temperature time constant,
$Q(T_p)$ is a function of $T_p$ accounting for particle temperature
source and/or heating and cooling of particles due to radiation,
e.g. see cases of optically thick and thin limits in situation
protoplanetary disk \cite{GL04}. The collective behavior of
particles can be represented by the Eulerian instantaneous equations
in physical space ${\bf x}$ and time $t$ for particles number
density $n({\bf x},t)$, velocity $V_i({\bf x},t)$ (obtained from Eq.
(\ref{l1}) \cite{PSC04}) and for temperature $\Theta$ which can be
formed from Eq. (\ref{l2}) in an usual manner, written as
\begin{equation}
\frac{\partial \Theta}{\partial t}+{V}_j\frac{\partial
\Theta}{\partial x_j}=\beta_\theta (T-\Theta)+Q(\Theta).
\label{eth1}
\end{equation}
To extract statistical information for collective behavior of
particles from the instantaneous equations, the density-weighted
average (denoted by overbar) of the equations is preferable instead
of non-weighted average due to the established reason
\cite{PSC04}. The mean number density $N=\langle n\rangle$ then
satisfies ${\partial N}/{\partial t}+{\partial (N\overline
{V}_{j})}/{\partial x_j}=0$, density-weighted averages
$\overline{V}_i$ and $\overline{\Theta}$ satisfy
\begin{equation}
D_t\overline{V}_j
+\frac{\partial \overline{v_n'v_j'}}{\partial x_n}=\beta_v
(\langle{U}_j\rangle-\overline{V}_j)+R_j +\beta_v \overline{u_j''},
\label{eqvb}
\end{equation}
\begin{equation}
D_t\overline{\Theta}
+\frac{\partial \overline{v_n'\theta'}}{\partial x_n}=\beta_\theta
(\langle{T}\rangle-\overline{\Theta})
+\overline{Q}-\overline{v_n'\theta'} \frac{\partial \ln N}{\partial
x_n}+\beta_\theta \overline{t''}\label{eqtemp}
\end{equation}
with
$R_j=2\epsilon_{jab}\overline{V}_a\Omega_b+{f}_j-\overline{v_n'v_j'}
{\partial \ln N}/{\partial x_n}$ and operator $D_t={\partial
}/{\partial t}+\overline{V}_i{\partial }/{\partial x_i}$. Also for
any instantaneous variable $A$, $\overline{A}=\langle n
A\rangle/N$ with ensemble average denoted by $\langle \,\rangle$,
$a'$ and $a''$ are fluctuations in $A=\overline{A}+a'=\langle A
\rangle +a''$ over $\overline {A}$ and $\langle A\rangle$,
respectively. In Eqs. (\ref{eqvb}) and (\ref{eqtemp}), $u_j''$ and
$t''$ represent respectively, fluctuations in fluid velocity and
temperature, in the vicinity of particle, over the fluid mean
velocity $\langle U_i\rangle$ and temperature $\langle T\rangle$.
The appeared unknown terms $\overline {v_n'v_j'},\overline
{v_n'\theta'}$ and, in particular,
$\overline{u_j''},\overline{t''}$ are difficult to model from the
instantaneous equations but can be obtained with ease in PDF
approach \cite{Reeks92,PM03,PSC04}.

\section{Non-Isothermal, Isotropic Turbulence Case}
Before obtaining
expressions for the unknown terms, consider Eqs. (\ref{eqvb}) and
(\ref{eqtemp}) for an ideal situation homogeneous, isotropic,
non-isothermal fluid turbulence with uniform $\langle {T}\rangle$,
$\Omega_b=0$ and having dispersed particles settling under the gravity
with $Q=0,f_i=-\delta_{i3}g$ where $g$ is acceleration due to gravity.
The equations simplify to $\overline{V}_3=-g/\beta_v+\overline{u_3''}$
and $\overline{\Theta}=\langle {T}\rangle+ \overline{t''}$ having
additional drift velocity $\overline{u_3''}$ and temperature
$\overline{t''}$, for the particles, caused by the turbulence velocity
and temperature structures. For the isotropic case, though
$\overline{u_3''}$ is nonzero \cite{Reeks01}, we show now that
$\overline{t''}=0$ when the effects of particle temperature on the
fluid temperature are neglected (i.e. one way coupling of
temperature). For the one-way temperature coupling and diffusivity
constant $\alpha$, $t''$ is governed by $\frac{\partial t''}{\partial
  t}+u_j''\frac{\partial t''}{\partial x_j}=\alpha \frac{\partial^2
  t''}{\partial x_j\partial x_j}$ and which suggests that for each
realization of $u_j''$, at any time $t$, $t''$ and $-t''$ are equally
probable at any location of the dispersed particle. This suggests that
the density-weighted average of $t''$ vanishes
i.e. $\overline{t''}=0$. All these imply that particles' mean
temperature $\overline{\Theta}=\langle {T}\rangle$ is identical in
both the cases of isotropic non-isothermal turbulent and stationary
fluids, both having identical uniform mean fluid temperature $\langle
{T}\rangle$.

\section{PDF Approach}
We use PDF approach to obtain expressions for
unknown terms and, in particular, $\overline{u_i''}$ and
$\overline{t''}$ in general situation of turbulent flows. The
equation for ensemble average of phase space density $W({\bf x},{\bf
v},\theta,t)$ of the particles, representing PDF equation, is
\begin{eqnarray}
\frac{\partial \langle W \rangle }{\partial t}+\frac{\partial
}{\partial x_i}v_i \langle W \rangle+\frac{\partial }{\partial v_i}
\beta_v  (\langle U_i \rangle -v_i)\langle W \rangle +
\frac{\partial}{\partial v_i}(2\epsilon_{iab}v_a
\Omega_b+f_i)\langle W \rangle\nonumber \\
 +\frac{\partial }{\partial \theta}[  \beta_\theta
(\langle T\rangle-\theta) +Q]\langle W\rangle= -\frac{\partial
}{\partial v_i}[  \beta_v  \langle u_i''W \rangle]-\frac{\partial
}{\partial \theta}[ \beta_\theta \langle t''W \rangle],\label{R3}
\end{eqnarray}
which can be obtained from Eqs. (\ref{l1}) and (\ref{l2}) and
Liouville's theorem \cite{PM03}. Here ${\bf x},{\bf v}$, and $\theta$
are phase space variables corresponding to ${\bf X},{\bf V}$, and
$T_p$, respectively. The expressions for unknown terms $\langle u_i''W
\rangle$ and $\langle t''W \rangle$ appearing in Eq. (\ref{R3}) can be
obtained by employing Furutsu-Donsker-Novikov functional formula
\cite{Zaichik99,Derevich00,HMR99b,PM03b}, which are {\it exact}
expressions when $u_i''$ and $t''$ along the particle path have
Gaussian distribution. The expressions are
\begin{equation}
{\beta_v\langle u_i''W\rangle}=-\left[\frac{\partial}{\partial
x_k} \lambda_{ki}+ \frac{\partial}{\partial v_k} \mu_{ki} +
\frac{\partial}{\partial \theta}\omega_i -\gamma_i\right]\langle W
\rangle, \label{uw2}
\end{equation}
\begin{equation}
{\beta_\theta\langle t''W\rangle}=-\left[\frac{\partial}{\partial
x_k} \Lambda_{k}+ \frac{\partial}{\partial v_k} \Pi_k +
\frac{\partial}{\partial \theta}\Omega -\Gamma\right]\langle W
\rangle. \label{tw2}
\end{equation}
where various tensors are
\begin{eqnarray}
 \Lambda_k&=&\beta_v\beta_\theta \int_0^tdt_2\langle t''
u_j''(t|t_2)\rangle G_{jk}(t_2|t), \label{lambda11} \\
\Pi_k&=&\beta_v\beta_\theta \int_0^tdt_2\langle t''
u_j''(t|t_2)\rangle \frac{d}{dt}{G}_{jk}(t_2|t), \label{pi123}\\
\Gamma &=&\beta_v \beta_\theta \int_0^tdt_2\left\langle
u_j''(t|t_2){\partial t''}/{\partial x_k} \right\rangle
G_{jk}(t_2|t) \label{gamma1} \\
\Omega&=&\beta_v \beta_\theta \int_0^tdt_2 \langle t''
u_j''(t|t_2)\rangle G_{j}(t_2|t)+\Omega_2
\label{omega1}
\end{eqnarray}
with $\Omega_2=\beta_\theta^2 \int_0^tdt_2\langle
t''t''(t|t_2)\rangle G^\theta(t_2|t)$ and expressions for
remaining tensors can be obtained by writing $\beta_v u_i''$
instead of $\beta_\theta t''$ on the right-hand side (rhs) of Eqs.
(\ref{lambda11})-(\ref{omega1}) and in $\Omega_2$ and changing $
\Lambda_k\rightarrow \lambda_{ki},
\Pi_k\rightarrow\mu_{ki},\Omega\rightarrow\omega_i$ and
$\Gamma\rightarrow\gamma_i$. In these expressions, shorthand
notations $u_i''(t|t_2), t''(t|t_2)$ represent $u_i''({\bf x},{\bf
v},\theta,t|t_2), t''({\bf x},{\bf v},\theta,t|t_2)$, the argument
$({\bf x},{\bf v},\theta,t|t_2)$ represents the value of $u_i''$
and $t''$ at time $t_2$ in the vicinity of particle that passes
through ${\bf x}$ at time $t$ with velocity ${\bf v}$ and
temperature $\theta$. Also, $u_i''$ and $t''$ represent
$u_i''({\bf x},t)$ and $t''({\bf x},t)$, respectively and now
onwards too. The equations for $G_{jk}(t_2|t), G_j(t_2|t),
G^\theta (t_2|t)\, \forall\,\,t \ge t_2$ are
\begin{equation}
{\mathcal{D}}G_{jk}(t_2|t)
-\beta_v G_{ji}\frac{\partial  \langle {U}_k\rangle}{\partial x_i}
-2\epsilon_{kab}\Omega_b \frac{dG_{ja}}{dt}-\frac{\partial
f_k}{\partial x_i}G_{ji} =\delta_{jk}\delta(t-t_2) \label{ngreen1}
\end{equation}
\begin{equation}
\frac{d}{dt}G_j(t_2|t)-\beta_\theta G_{jk}\frac{\partial \langle
{T}\rangle}{\partial x_k}-\frac{\partial Q(\theta)}{\partial
\theta}G_{j}+\beta_\theta G_j=0 \label{ngreen2}
\end{equation}
\begin{equation}
\frac{d}{dt}G^\theta (t_2|t)+\beta_\theta G^\theta-\frac{\partial
Q(\theta) }{\partial \theta}G^\theta=\delta(t-t_2) \label{ngreen3}
\end{equation}
where operator ${\mathcal{D}}={d^2}/{dt^2}+\beta_v {d}/{dt}$. For
later convenience, $\Lambda_i$ and $\Gamma$ are written in
different forms as
\begin{equation}
\frac{\Lambda_{i}}{\beta_\theta}=\int_0^tds\langle t''\Delta v_i
\rangle, \,\, \frac{\Gamma}{\beta_\theta}=\int_0^tds\langle \Delta
v_i{\partial t''}/{\partial x_i}\rangle \label{lga2}
\end{equation}
where $\Delta v_i=\int_0^sdt_2\,\beta_v
u_k''(t|t_2)\frac{dG_{ki}(t_2|s)}{ds}$ represents change in particle
velocity due to $\beta_v u_k''(t|t_2)$, during time $0$ to $s$,
along the trajectory that passes through ${\bf x}, {\bf v}$ at $t$.

The various tensors contain statistical properties related to
$u_i''$ and $t''$ along the particle path and which can be
obtained from the modelled equation for $u_i''$ along the particle
path \cite{PSC04}, exact equation for $t''$
\begin{equation}
\frac{dt''}{dt}=\alpha\nabla^2 t''-u_i''\frac{\partial \langle
T\rangle}{\partial x_i}+\langle u_i''\frac{\partial t''}{\partial
x_i}\rangle-(U_i-V_i)\frac{\partial t''}{\partial x_i} \label{lagt}
\end{equation}
with model $\alpha \nabla^2 t''=-t''/\tilde{T}_\theta$ based on
the simple IEM model \cite{Dopazo94} suggested for single-phase
turbulence. It should be noted that integral time scale
$\tilde{T}_\theta$ is different than the time scale of IEM model.
To obtain $\langle t''u_j''(t|t_2)\rangle$, multiply Eq.
(\ref{lagt}) by $u_j''(t|t_2)$ and take ensemble average to yield
${d\langle t''u_j''(t|t_2)\rangle}/{dt}=-{\langle
t''u_j''(t|t_2)\rangle}/{\tilde{T}_\theta}+S_j$ with $S_j=-\langle
u_j''(t|t_2)[u_i''\frac{\partial \langle T\rangle}{\partial
x_i}+(U_i-V_i)\frac{\partial t''}{\partial x_i}]\rangle$. Further,
the first order solution for $\langle t''u_j''(t|t_2)\rangle$ can
be obtained by neglecting $S_j$ and the result is an exponential
form $\langle t''u_j''(t|t_2)\rangle=\langle
t''(t|t_2)u_j''(t|t_2)\rangle e^{-(t-t_2)/\tilde{T}_\theta}$ which
is considered here for further analysis.

The continuum equations for the dispersed phase can be obtained by
taking various moments of Eq. (\ref{R3}) after substituting for
$\beta_v\langle u_i'W\rangle$ and $\beta_\theta\langle t'W\rangle$
from Eqs. (\ref{uw2}) and (\ref{tw2}). The equations for
$\overline{V}_i=\frac{1}{N}\int v_i\langle W \rangle d{\bf
v}d\theta$ and $\overline{\Theta}=\frac{1}{N}\int \theta\langle W
\rangle d{\bf v}d\theta$ are identical to Eqs. (\ref{eqvb}) and
(\ref{eqtemp}), respectively, with
\begin{equation}
\beta_v
\overline{u_j''}=-\frac{\overline{\lambda}_{ij}}{N}\frac{\partial
N }{\partial x_i}-\frac{\partial}{\partial
x_i}\overline{\lambda}_{ij}+\overline{\gamma}_j,\label{ueqt}
\end{equation}
\begin{equation}
\beta_\theta\overline{t''}=-\frac{\overline{\Lambda}_k}{N}\frac{\partial
N}{\partial x_k} -\frac{\partial \overline{\Lambda}_i}{\partial
x_i}+\overline{\Gamma}.\label{teqt}
\end{equation}
Though the equations for higher order correlations
$\overline{v_n'v_j'},\overline{v_n'\theta'}$ etc.
can also be derived from the PDF equation, we focus our discussion
on different phenomena contained in Eq. (\ref{teqt}) for particles'
temperature.


\section{New Effects}
The phenomena contained in various terms
of Eq. (\ref{ueqt}) for additional drift of particles are
described earlier \cite{PSC04}. Here we discuss phenomena
related to temperature of collective particles as described by Eq.
(\ref{teqt}). For the discussion, we consider density weighted
average of various tensors to be equal to their respective
instantaneous values, {\it e.g.}
$\overline{\lambda}_{ij}=\lambda_{ij}$. It should be noted that
the form of Eqs. (\ref{eqtemp}) and (\ref{teqt}) remains unchanged
irrespective of whether the rotation and gravitation exist or not.
The rotation and gravitation effects are felt by these equations
through $G_{ij}$ and $G_j$ appearing in various tensors. The first
term on the rhs of Eq. (\ref{teqt}) represents phenomenon of
additional increase or decrease in mean temperature of particles
caused by combination of statistical nature of the turbulence
structure represented by $\Lambda_k$ and gradient of particles
mean number density in general situation of non-isothermal
two-phase turbulent flows. The rhs of Eq. (\ref{teqt}) vanishes
when fluid variables along the particle path are correlated by
delta function in time as suggested by expressions for $\Lambda_k$
and $\Gamma$ along with $G_{jk}(t|t)=0$ \cite{PSC04}. When the
correlation time is finite, the last two terms of Eq. (\ref{teqt})
account for certain phenomena which we discuss now for cases of
(1) slow rotation i.e. $\Omega_b$ is small and (2) fast rotation
without gravity i.e. $f^g_i=0$.

\section{Slow Rotation Case}
For slow rotation, we simplify the last two terms in Eq.
(\ref{teqt}) by expanding the solution of Eq. (\ref{ngreen1}) as
$G_{jk}=G^0_{jk}+G^1_{jk}$ with
$G^0_{jk}=G^0(t_2|t)\delta_{jk}=\delta_{jk}[1-e^{-\beta_v(t-t_2)}]/\beta_v$
and
\begin{equation}
{\mathcal{D}}G^1_{jk} -\beta_v G^0_{ji}\frac{\partial \langle
{U}_k\rangle}{\partial x_i} -2\epsilon_{kab}\Omega_b
\frac{dG^0_{ja}}{dt}-\frac{\partial f_k}{\partial x_i}G^0_{ji}=0,
\label{g02}
\end{equation}
using exponential form for $\langle t''u_j''(t|t_2)\rangle$ and
$\langle t''(t|t_2)u_j''(t|t_2)\rangle\cong\langle
t''u_j''\rangle$. The simplification yields
\begin{eqnarray}
-\frac{\partial}{\partial x_i}{\Lambda}_i+{\Gamma}&=&
-\beta_v\beta_\theta\int_0^tdt_1 \langle t''u_k''(t|t_1) \rangle
\frac{\partial G_{ki}(t_1|t)}{\partial x_i }
-\beta_v\beta_\theta\int_0^tdt_1 G_{ki}^1(t_1|t)
e^{-(t-t_1)/\tilde{T}_{\theta}}
\langle t''{\partial u_k''}/{\partial x_i } \rangle \nonumber \\
&-&\langle t''{\partial u_k''}/{\partial x_k }\rangle \beta_\theta
A
\label{ther2}
\end{eqnarray}
where $A=\{
\tilde{T}_\theta[1-e^{-t/\tilde{T}_\theta}]+\frac{\tilde{T}_\theta}{\beta_v
\tilde{T}_\theta+1}[e^{-t(\beta_v+1/\tilde{T}_\theta)}-1]\}$. Using
the solution for $G_{ki}^1$ from Eq. (\ref{g02}), the second term on
the rhs of Eq. (\ref{ther2}) can be further simplified as
\begin{equation}
-\left\langle\frac{\partial u_k''}{\partial x_i } t'' \right\rangle
[\beta_v^2\beta_\theta \frac{\partial
 \langle U_i\rangle}{\partial x_k} A_1'(t)
 -2\beta_v\beta_\theta
\epsilon_{ikb}\Omega_b  A_2'(t)-\beta_v\beta_\theta\frac{\partial
(f^c_i+f^g_i)}{\partial x_k} A_1'(t)], \label{phen2}
\end{equation}
where $ A_1'=\int_0^tdt_1
[e^{-\frac{(t-t_1)}{\tilde{T}_{\theta}}}\int_{t_1}^tdt_2G^0(t_2|t)G^0(t_1|t_2)]$
and $
 A_2'=\int_0^tdt_1
[e^{-\frac{(t-t_1)}{\tilde{T}_{\theta}}}\int_{t_1}^tdt_2G^0(t_2|t)\frac{d}{dt_2}G^0(t_1|t_2)]
$. The first term in (\ref{phen2}) represents contribution to
heating/cooling of particles due to mean shear rate. The second term
in (\ref{phen2}) suggests that correlation between $t''$ and the
component of fluid vorticity at particle location
($\epsilon_{bik}\frac{\partial U_k({\bf x},t)}{\partial x_i }$) in
the direction of rotation $\Omega_b$ produces additional
heating/cooling for particle phase and whose origin is the Coriolis
force. The last term in (\ref{phen2}) represents heating/cooling of
the dispersed phase due to the interaction of centrifugal and
gravitational forces with turbulence fluctuations $u_i''$ and $t''$.
For further discussion, we consider a finite value for rotation only
about $x_3$ axis {\it i.e.} $\Omega_1=\Omega_2=0$ and $\Omega_3\neq
0$. Then, the centrifugal part of $f_k$ is
$f^c_k=x_k\Omega_3^2-\delta_{k3}x_k\Omega_3^2$ and its contribution
to the last term in (\ref{phen2}) becomes
\begin{equation}
-\beta_v \beta_\theta A_1'(t)\Omega_3^2\langle t''\{{\partial
u_k''}/{\partial x_k} -{\partial u_3''}/{\partial x_3}\}\rangle,
\label{centrigra}
\end{equation}
which describes the phenomenon of heating/cooling of the dispersed
phase in the presence of a centrifugal force. The effects of
gravitational force contained in the last term in (\ref{phen2}) can
be simplified as
\begin{equation}
-\beta_v\beta_\theta A_1'(t)[\langle t''{\partial u_k''}/{\partial
x_k } \rangle  f^g + \langle t''{\partial u_k''}/{\partial x_i }
\rangle x_i\frac{\partial f^g}{\partial x_k}]\label{grav2}
\end{equation}
where the last term arises due to variation of gravitational
acceleration $f^g$ in space.

\section{Fast rotation without gravity}
For large value of
$t$, Eqs. (\ref{lga2}) suggest that most contributions to
$\Lambda_i$ and $\Gamma$ come from the correlations of $t''$ and
$\partial t''/\partial x_i$ with velocity of particles, between
$s(<t)$ and $t$, present in the region near to ${\bf x}$ and $t$
and passing through ${\bf x}$ at $t$. Assuming exponential form
for these correlations with integral time scale
$\tilde{T}_\theta$, the last two terms in Eq. (\ref{teqt}) become
approximately equal to $-\beta_\theta \langle t''{\partial
V_i({\bf x},t)}/{\partial x_i}\rangle
\int_0^tds\,e^{\frac{s-t}{\tilde{T}_\theta}}$ and which, along
with an approximate relation $\frac{\partial V_i({\bf
x},t)}{\partial
x_i}\approx\frac{2\tau_p\Omega_i\epsilon_{iab}\frac{\partial U_b
}{\partial x_a}}{1+4\tau_p^2 \Omega_k\Omega_k}$ for fast rotation
without gravity \cite{EKR98b}], suggests phenomenon of additional
cooling/heating of particles caused by fast rotation. This
phenomenon does not diverge with the growth of the angular
velocity.

\section{Effect of Fluid Compressibility}
Fluid compressibility gives rise to interesting effects, of
turbulent thermal diffusion and barrodiffusion of particles, which
are contained in Eq. (\ref{ueqt}) \cite{PM02b}. For particles
dispersed in a compressible ideal gas, compressibility effects on
the temperature of the particles can be obtained by using the
following relation \cite{PM03}
\begin{equation} \left\langle
t''{\partial u_k''}/{\partial x_k}\right\rangle\cong \langle
t''u_k''\rangle  [  {\partial \ln\langle T\rangle}/{\partial x_k}-
{\partial \ln\langle P \rangle}/{\partial x_k}]. \label{uiuit}
\end{equation}
Substituting it into the last term in Eq. (\ref{ther2}) suggests
turbulent thermal and pressure effects on the heating and cooling
of the dispersed phase analogous to the two phenomena of turbulent
thermal diffusion and barodiffusion \cite{PM02b}. Substituting
Eq. (\ref{uiuit}) into Eqs. (\ref{centrigra}) and (\ref{grav2})
suggests that the first term on the rhs of these equations further
adds to these thermal and pressure effects in the presence of
rotation and gravitation field.

\section{Concluding Remarks}
It has been our goal to accurately describe and discover the
phenomena related to important situation of dispersed phase of
non-isothermal particles in rotating turbulent flows under the
influence of gravitational field. The mean temperature
$\overline{\Theta}$ of the particle phase is found to be affected by
the term $\overline{t''}$ in Eq. (\ref{eqtemp}) containing new
effects caused by mutual interactions of turbulent structure of
fluid temperature and velocity with the rotation and gravitation
field. These effects and their accurate description have been
surfaced through the application of mathematical rigorous PDF
approach having established capability in capturing certain
phenomena (see \cite{PSC04} and references cited therein).


\begin{thebibliography}{10}

\bibitem{RY89}
R.~R. Rogers and M.~K. Yau, {\em A Short Course in Cloud Physics}
  (Butterworth-{Heinemann}, Woburn, MA, 1989).

\bibitem{Friedlander00}
S.~K. Friedlander, {\em Smoke, Dust and Haze. Fundamental of Aerosol Dynamics}
  (Oxford University Press, New York, NY, 2000).

\bibitem{Bannon02}
P.~R. Bannon, Journal of the Atmospheric Sciences {\bf 59},  1967  (2002).

\bibitem{GL04}
P. Garaud and D.~N.~C. Lin, The Astrophysical Journal {\bf 608},  1050  (2004).

\bibitem{KRGB04}
V.~M. Khazins, V.~A. Rybakov, R. Greeley, and M. Balme, Solar System Research
  {\bf 38},  12  (2004).

\bibitem{CTTV75}
M. Caporaloni, F. Tampieri, F. Trombetti, and O. Vittori, J. Atmos. Sci. {\bf
  32},  565  (1975).

\bibitem{EF94}
J.~K. Eaton and J.~R. Fessler, Int. J. Multiphase Flow Suppl. {\bf 20},  169
  (1994).

\bibitem{EKR98}
T. Elperin, N. Kleeorin, and I. Rogachevskii, Phys. Rev. E {\bf 58},  3113
  (1998).

\bibitem{BFF01}
E. Balkovsky, G. Falkovich, and A. Fouxon, Phys. Rev. Lett. {\bf 86},  2790
  (2001).

\bibitem{PM02b}
R.~V.~R. Pandya and F. Mashayek, Phys. Rev. Lett. {\bf 88},  044501  (2002).

\bibitem{PSC04}
R.~V.~R. Pandya, P. Stansell, and J. Cosgrove, Physical Review E {\bf 70},
  025301(R)  (2004).

\bibitem{Boltzmann64}
L. Boltzmann, {\em Lectures on Gas Theory} (University of California Press,
  Berkeley and Los Angeles, 1964), translated by Stephen G. Brush.

\bibitem{McComb90}
W.~D. McComb, {\em The Physics of Fluid Turbulence} (Oxford University Press,
  New York, NY, 1990).

\bibitem{PM03}
R.~V.~R. Pandya and F. Mashayek, J. Fluid Mech. {\bf 475},  205  (2003).

\bibitem{DZ90}
I.~V. Derevich and L.~I. Zaichik, Journal of Applied Mathematics and Mechanics
  (Prikl. Mat. Mekh.) {\bf 54},  631  (1990).

\bibitem{Reeks91}
M.~W. Reeks, Phys. Fluids {\bf 3},  446  (1991).

\bibitem{Reeks92}
M.~W. Reeks, Phys. Fluids {\bf 4},  1290  (1992).

\bibitem{Zaichik99}
L.~I. Zaichik, Phys. Fluids {\bf 11},  1521  (1999).

\bibitem{PM99}
J. Pozorski and J.~P. Minier, Phys. Rev. E {\bf 59},  855  (1999).

\bibitem{Derevich00}
I.~V. Derevich, Int. J. Heat Mass Transfer {\bf 43},  3709  (2000).

\bibitem{MP01}
J.-P. Minier and E. Peirano, Physics Reports {\bf 352},  1  (2001).

\bibitem{Reeks01}
M.~W. Reeks, {\em Paper 187, presented at 4th International Conference on
  Multiphase Flow, May 27-June 1} (PUBLISHER, New Orleans, LA, USA, 2001).

\bibitem{HMR99b}
K.~E. Hyland, S. McKee, and M.~W. Reeks, J. Phys. A: Math. Gen. {\bf 32},  6169
   (1999).

\bibitem{PM03b}
R.~V.~R. Pandya and F. Mashayek, AIAA Journal {\bf 41},  841  (2003).

\bibitem{Dopazo94}
C. Dopazo,  in {\em Turbulent Reacting Flows}, edited by P.~A. Libby and F.~A.
  Williams (Academic Press, London, UK, 1994), Chap.~7, pp.\ 375--474.

\bibitem{EKR98b}
T. Elperin, N. Kleeorin, and I. Rogachevskii, Physical Review Letters {\bf 81},
   2898  (1998).

\end{thebibliography}
\bibliographystyle{prsty}

\end{document}